\documentclass[a4paper,preprintnumbers,nofootinbib,twocolumn]{revtex4}
\pdfoutput=1
\usepackage{amsmath,bbm,latexsym,url,graphicx}

\def\lsim{\mathrel{\raise.3ex\hbox{$<$\kern-.75em\lower1ex\hbox{$\sim$}}}}
\def\gsim{\mathrel{\raise.3ex\hbox{$>$\kern-.75em\lower1ex\hbox{$\sim$}}}}
\def\gev{\mbox{GeV/c${}^2$}}

\begin{document}

%%%\vphantom{2cm}
\title{\vphantom{5cm} The Higgs mass coincidence problem: why is the higgs mass $m_H^2=m_Z m_t$?}
%\title{\vphantom{5cm} The Higgs and top masses: why is the higgs mass $m_H^2=m_Z m_t$?}

\date{August 12,2012}

%\preprint{IFT--12-45}
\preprint{FISPAC-12/132}
\preprint{UQBAR-TH-12/718}

\author{E. Torrente-Lujan}
%\email[]{torrente@cern.ch}
\email[]{etl@um.es}
\affiliation{Dept. Physics,  Murcia U.,  SPAIN}

\begin{abstract}
On the light of the recent LHC boson discovery,
we present a phenomenological evaluation of the  ratio
$\rho_t=m_Z m_t/m_H^2$, from the LHC combined $m_H$  value, 
we get ($ (1\sigma)$)
$$\rho_t^{(exp)}= 0.9956\pm 0.0081.$$
This value  is close to one with a precision of the order 
$\sim 1\%$.
%, not far from the precision $\rho=m_W^2/m_Z^2\cos^2\theta_W$ is presentely known.
%
Similarly we evaluate the ratio
 $\rho_{Wt}=(m_W + m_t)/(2 m_H)$. From the up-to-date 
mass values we get $\rho_{Wt}^{(exp)}= 1.0066\pm 0.0035\; (1\sigma).$
%This is close to one but  the favoured value is around 2$\sigma$ from being exactly one.
The Higgs mass is numerically close (at the $1\%$ level) 
to the $m_H\sim (m_W+m_t)/2$.
%
%The   relations $\rho_t\simeq\rho_{Wt}\simeq 1$ imply that any two of the quantities $m_H,m_W,m_Z,m_t$ can be written in terms of the other two. 
%
From these relations we can write any two mass 
ratios as a function of, exclusively, the Weinberg angle 
(with a precision of the order of $1\%$ or better):
\begin{eqnarray}
\frac{m_i}{m_j}&\simeq & f_{ij}(\theta_W),\quad i,j=W,Z,H,t.
\end{eqnarray}
For example:
$m_H/m_Z \simeq 1+\sqrt{2} s_{\theta_W/2}^2$,
$m_H/m_t c_{\theta_W} \simeq 1-\sqrt{2}s_{ \theta_W/2}^2$.
%\begin{eqnarray}
%\frac{m_W}{m_Z} &\simeq& \cos\theta_W ,\\
%\frac{m_H}{m_Z} &\simeq& 1+\sqrt{2}\sin^2 \frac{\theta_W}{2} ,\\
%\frac{m_H}{m_t}\cos\theta_W &\simeq& 1-\sqrt{2}\sin^2 \frac{\theta_W}{2}.
%\end{eqnarray}
In the limit $\cos\theta_W\to 1$ all the masses 
would become equal $m_Z=m_W=m_t=m_H$.
%It is tempting to think that such a value, it is not a mere coincidence but, on naturalness grounds, a signal of some more deeper  symmetry. In a model independent way, $\rho_t$ can be viewed as the ratio of the highest massive representatives  of the spin $(0,1/2,1)$ SM  and, to a very good precision  the LHC evidence tell us that $ m_{s=1} m_{s=1/2} /m_{s=0}^2 \simeq  1.$ Somehow  the ``lowest'' scalar particle mass  is the geometric mean of the highest  spin 1, 1/2 masses.
%

We review the theoretical situation of this ratio in the SM and 
beyond. 
In the SM  these relations are rather  stable under 
RGE pointing out to some underlying UV symmetry. 
In the SM such a ratio hints for a non-casual relation
of the type 
$\lambda \simeq \kappa \left (g^2+{g'}{}^2\right )$
with $\kappa\simeq 1+o(g/g_t)$.
Moreover the existence of relations 
$m_i/m_j \simeq  f_{ij}(\theta_W)$ could be interpreted as a 
hint for a role of the $SU(2)_c$ custodial symmetry, 
together with other unknown mechanism.
Without a symmetry at hand to explain then in the SM, 
it arises a Higgs mass coincidence problem, why 
the ratios  $\rho_t,\rho_{Wt}$ are so close to one, can we find
a mechanism that naturally  gives 
$m_H^2=m_Z m_t$, $2m_H= m_W+m_t$?. 

%%\vspace{3cm}

\end{abstract}

%\newpage

\maketitle

%%%%%%%%%%%%%%%%%%%%%%%%%%%%%%%%%%%%%%%%%%%%%%%%%%%%%%%%%%%%%%%%%%
\section{The ratio $\rho_t=m_Z m_t/m_H$}

\noindent
%%{\bf I.1.}
The problem of the mass of elementary particles 
has two independent aspects in particle 
physics. The first, how mass arises, it is answered in the SM 
by the Higgs mechanism. 
\footnote{For composite particles, i.e. hadrons, 
the dynamical generation of mass is indeed a dominant 
mechanism.}
The second aspect is  why different  elementary 
particles have their specific  masses. 
Unless electromagnetic charge, there is no any, exact or 
approximate, known relation, structure or hierarchy among 
the masses of the SM elementary particles.

Evidence in favour of the existence of 
a boson with spin $s=0$ and properties compatible with 
those of a SM Higgs particle 
(production cross sections, mass-proportional couplings)
 has been presented by the LHC 
experiments \cite{atlas,cms}. 

\begin{figure}
\begin{tabular}{c}
\includegraphics[scale=0.5]{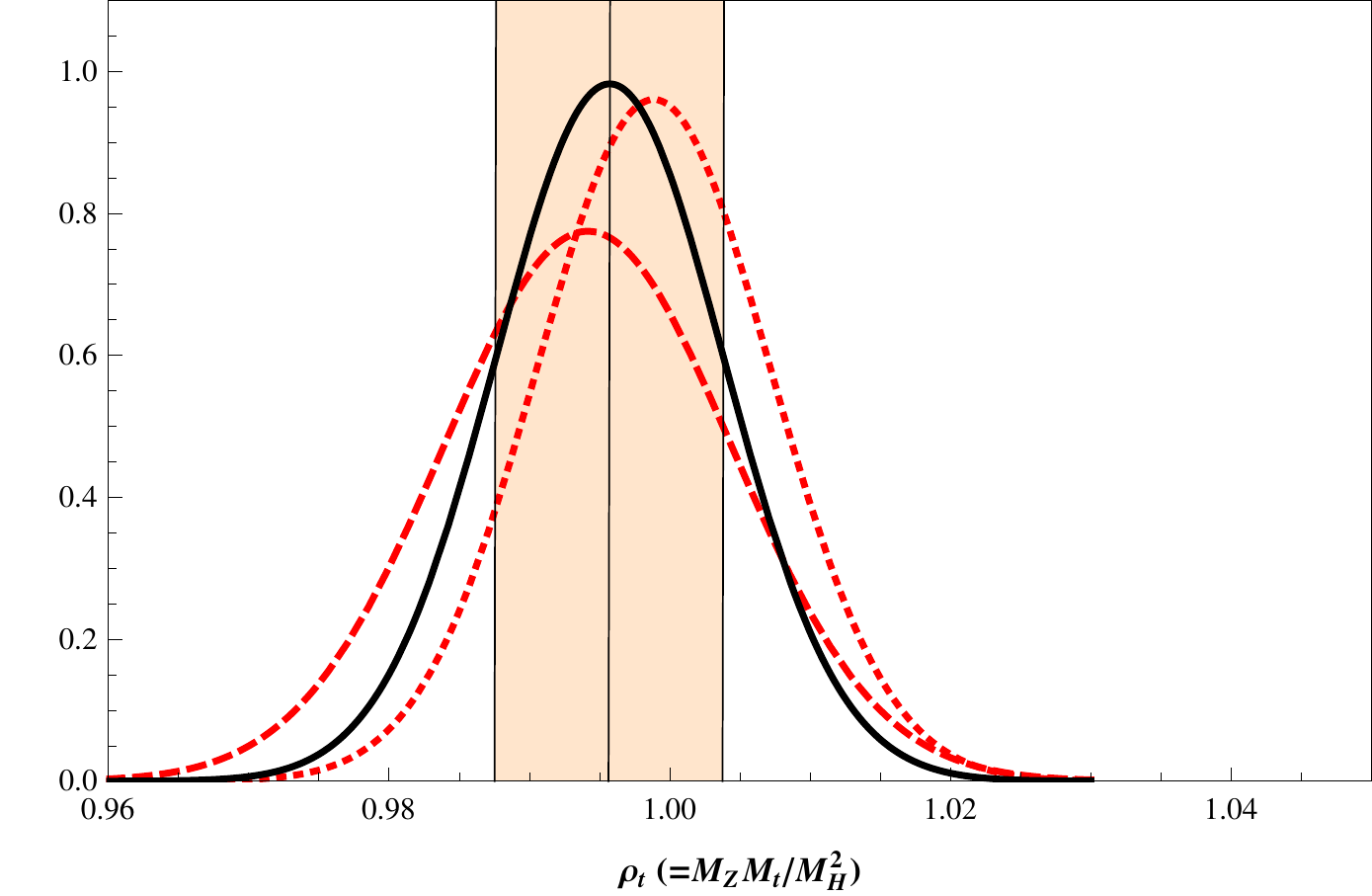}\\
\includegraphics[scale=0.5]{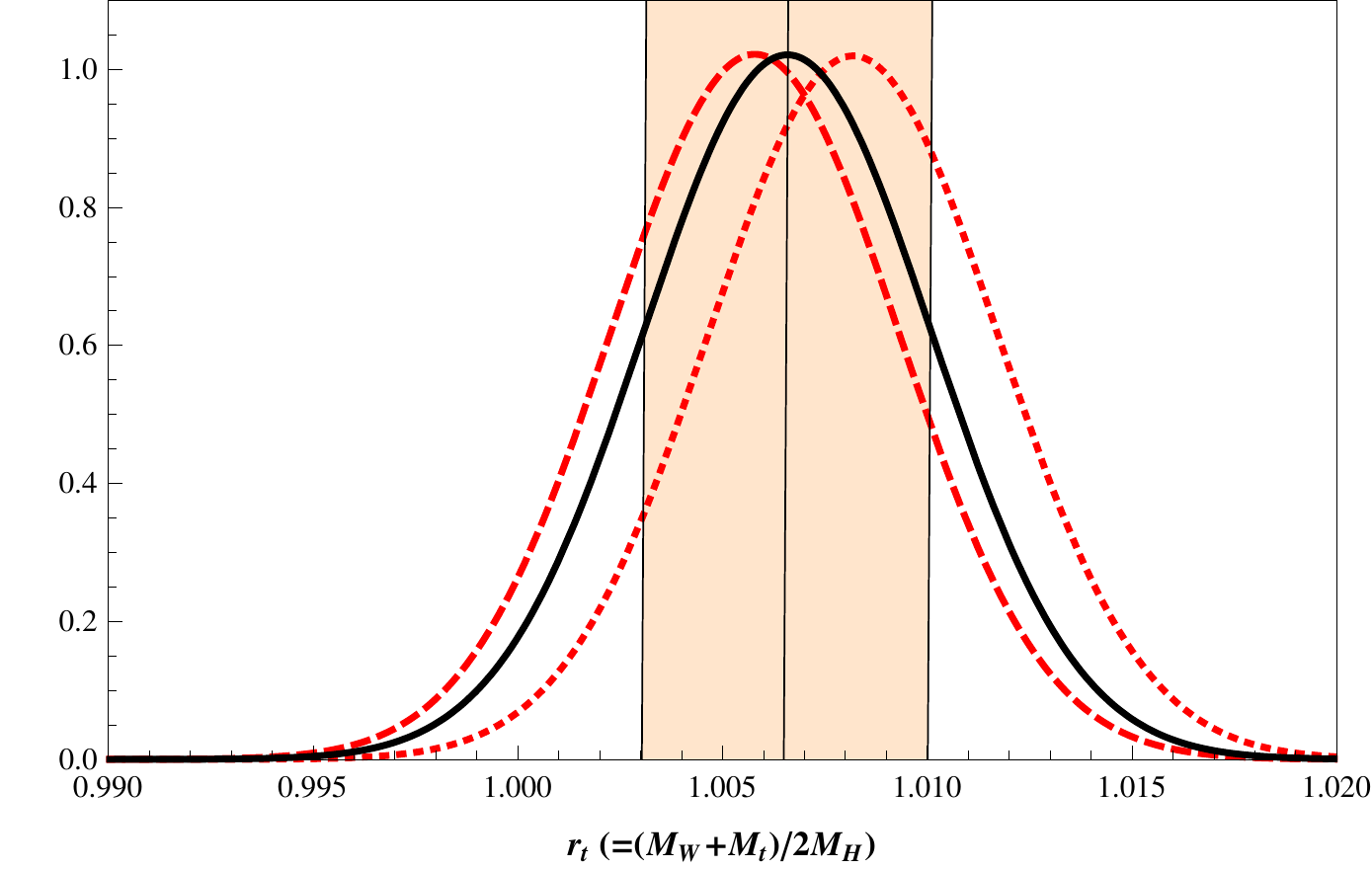}
\end{tabular}
\caption{MC generated Likelihood distributions 
for each of the quantities $\rho_t$ (up) and $\rho_{Wt}$ (down).
 The curves are in arbitrary units with normalized areas.
The curves correspond to the global averaged Higgs mass (continuos) and individual ATLAS and CMS values (dashed and dotted lines). The shaded area is the $1\sigma$ allowed region for each 
of the parameters.}
\label{fig1}
\end{figure}

The aim of this work is to present some detailed 
phenomenological estimations of some ratios involving 
the experimental 
masses of the Higgs boson, the vector bosons and the top quark, 
the derivation  from them of some model independent 
expressions and detailed study of them in the framework 
of the SM.
 In the light of the recent results from the LHC coming
from the experiments ATLAS and CMS,
the parameter defined by the relation
\begin{eqnarray}
\rho_t &=& \frac{m_Z m_t}{m_H^2}
\end{eqnarray}
where $m_Z,m_t$ are the masses  of the $Z^0$ gauge boson 
and the top quark and $m_H$ is the Higgs mass has become  
experimentally measurable.
We estimate its current value
to be
\begin{eqnarray}
\rho_t^{(exp)} &=& 0.9956\pm 0.0081
\label{eq2}
\end{eqnarray}
where we have used the current values for \cite{pdg2012} 
\begin{eqnarray}
m_Z&=&91.1876\pm 0.0021 \quad \gev,\\
m_t&=&173.07\pm 0.52\pm 0.72 \quad \gev
\end{eqnarray}
and the combined value of the boson masses presented 
by ATLAS and CMS \cite{atlas,cms}, 
\begin{eqnarray}
m_H&=& 125.9\pm 0.4 \pm 0.4 \quad\gev.
\label{eq12zz}
\end{eqnarray}
The combined value of the boson mass 
is obtained by standard statistic techniques, we
 neglect correlations among the systematic component of the 
errors.
The value (\ref{eq2}) is obtained by a MC simulation. 
First, a distribution of the quotient is obtained by generating 
 Gaussian ensembles of the individual masses.
 Second,  symmetric gaussian fit, fig.(\ref{fig1}), is 
performed to the central part of this (close to symmetric) 
distribution avoiding the non-gaussian tail.
The central value and $1\sigma$ errors appearing 
in (\ref{eq2}) are extracted from this fitted gaussian.
If the individual values for each ot the experiments are used 
instead (using a similar statistical procedure), we get 
(see fig.(\ref{fig1})(up))
\begin{eqnarray}
\rho_t^{(exp)} &=& 0.9940\pm 0.0102 \quad\quad (m_{h,ATLAS}),\\
\rho_t^{(exp)} &=& 0.9990\pm 0.0085 \quad\quad (m_{h,CMS})
\end{eqnarray}
for boson masses respectively 
$$m_H=125.8\pm 0.4 \pm 0.4 \quad\gev$$ 
and 
$$m_H=126.0\pm 0.4 \pm 0.4 \quad\gev.$$
The conclusion is that the experimental value of the 
ratio $\rho_t$ is close to one with a precision of the order 
or less than $1\%$. This precision is not far from the precision at which the well known ratio
$$\rho=m_W^2/m_Z^2\cos^2\theta_W$$
is presentely measured, $\rho=1.0008\pm 0.001$ \cite{pdg2012} 
with $\theta_W$ the Weinberg angle and $m_W$ the charged 
electroweak gauge boson mass. 
The closeness of this parameter $\rho_t$ to one might 
be merely a coincidence which 
will dissapear with any new measurement or might be not.

Note that the ratio would be exactly one for a boson mass 
(and  nominal $m_Z,m_t$ PDG masses) of 
\begin{eqnarray}
m_H (\rho_t=1) &\simeq &125.6 \quad \gev,
\end{eqnarray}
a value somewhere inbetween of the $125-126$ range  of values 
currently measured by LHC and just on the borderline of the 
SM vacuum stability limits \cite{anton}.
 
The ratio $\rho_t$ would be still close to one, with a 
precision of  $5\%$, 
if the Higgs mass appear finally anywhere in the 
range $m_H=123-129$ \gev.
If we vary the top mass in the range $m_t\sim 170-175$ \gev similar 
results are obtained.

Similar ratios involving other  fermion masses instead of the 
top mass could be obviously defined, for example including 
all the fermions we could define $\rho_\Sigma$ as
\begin{eqnarray}
\rho_\Sigma &=& \frac{m_Z m_\Sigma}{m_H^2},% m_\Sigma^2=\sum_f m_f^2
\end{eqnarray}
with
\begin{eqnarray}
 m_\Sigma^2&=&\sum_f m_f^2
\end{eqnarray}
or including the third family quark doublet ($m_Q^2=m_t^2+m_b^2$)
we could define the ratio
\begin{eqnarray}
\rho_T &\equiv& \frac{m_Z m_Q}{m_H^2}, \\
&\simeq& \rho_t \left( 1+ 2 \left(\frac{m_b}{m_t}\right)^2 \right ).
\end{eqnarray}
In any case, any of these or similar ratios are deviated from $\rho_t$ by 
a very moderate quantity $(m_b/m_t)^2\simeq 10^{-3}$.

It is also interesting to consider 
an alternative way to express the closeness of the  ratio $\rho_t$
to one.
If we consider the individual mass rations $m_Z/m_H, m_H/m_t$, their 
current experimental values are 
%\footnote{ ``God'' or  ``golden'' particle?. The difference between any of the $m_H/m_Z$, $m_Z/m_H$ ratios and the Golden Ratio $(\surd 5+1)/2$ is a ``mere'' $15\%$. Equality would be exact if $m_t=m_H+m_Z$}.
\begin{eqnarray}
\frac{m_Z}{m_H} &=&0.725\pm 0.003, \label{eq12}\\
\frac{m_H}{m_t} &=&0.727\pm 0.005   \label{eq13}
\end{eqnarray}
where we have taken the LHC combined value of $m_H$. 
and  PDG $m_Z,m_t$ masses.
Both ratios are the same at the level of 1\% 
(and totally compatible at even higher precision according to 
present error bars).
Very similar results are obtained if we use any of the ATLAS or CMS individual measurements

Similarly to $\rho_t$ we define now another ratio of masses 
involving the Higgs, vector bosons and top quarks, 
whose experimental value is also seen to be close to one. 
Let us take 
\begin{eqnarray}
\rho_{Wt}&=& \frac{m_W+m_t}{2 m_H}
\end{eqnarray}
where $m_W$ is the mass of the $W$ boson.
We estimate the current value of this ratio (using 
a similar MC technique as explained above, see fig.(\ref{fig1})(down)) to be
\begin{eqnarray}
\rho_{Wt}^{(exp)}&=& 1.0066\pm 0.0035
\label{eq11zz}
\end{eqnarray}
where we have used the current value for $M_W$ \cite{pdg2012}
$$ M_W=80.385\pm 0.015 \gev$$
and the combined value for the Higgs mass, Eq.(\ref{eq12zz}).
If the individual values for each of the experiments are 
used instead, we get
\begin{eqnarray}
\rho_{Wt}^{(exp)} &=& 1.0082\pm 0.0036 \quad\quad (m_{h,ATLAS}),\\
\rho_{Wt}^{(exp)} &=& 1.0056\pm 0.0036 \quad\quad (m_{h,CMS}).
\end{eqnarray}
The experimental value of the ratio $\rho_{Wt}$ is close to 
one with a precision of the order of $1\%$. Nevertheless 
the favoured value is around 2$\sigma$ from being exactly one.
The Higgs mass is numerically close to the aritmethic average
of the charged boson and top masses $m_H\sim (m_W+m_t)/2$.
The ratios $\rho_t$ and $\rho_{Wt}$ are a priori independent.
%There is no  reason why the value of one of them should be close to one even if the other happens to be close to one.

The   relations $\rho_t\simeq\rho_{Wt}\simeq 1$ imply that 
any two of the quantities $m_H,m_W,m_Z,m_t$ can be written
in terms of the other two. Taking into account also 
the relation $\rho\simeq 1$ we can write any two mass 
ratios as a function of, exclusively, the Weinberg angle 
(with a precision of the order of $1\%$ or better):
\begin{eqnarray}
\frac{m_i}{m_j}&\simeq & f_{ij}(\theta_W),\quad i,j=W,Z,H,t.
\label{eq44zz}
\end{eqnarray}
Examples of these relations are:
\begin{eqnarray}
\frac{m_W}{m_Z} &\simeq& \cos\theta_W ,\\
\frac{m_H}{m_Z} &\simeq& 1+\sqrt{2}\sin^2 \frac{\theta_W}{2} ,\\
\frac{m_H}{m_t}\cos\theta_W &\simeq& 1-\sqrt{2}\sin^2 \frac{\theta_W}{2}.
\end{eqnarray}
In the limit $\cos\theta_W\to 1$ all the masses 
would become equal $m_Z=m_W=m_t=m_H$.

%%%%%%%%%%%%%%%%%%%%%%%%%%%%%%%%%%%%%%
%%%%%%%%%%%%%%%%%%%%%%%%%%%%%%%%%%%%%%%%%%%%

\section{In the SM}
\vspace{0.3cm}
%%{\bf 2.} 

The latest LHC measurements \cite{atlas,cms}, 
in particular the characteristics of the seen 
$H\to\gamma\gamma$ channel,
point to a preferred discovery of 
a neutral boson of spin 0.
In a model independent way, thus the quantity $\rho_t$ can 
be viewed as the ratio of the highest 
massive representatives  of the spin $(0,1/2,1)$ particles of 
the Standard Model and, to a very good precision 
 the experimental evidence tell us that
\begin{eqnarray}
\rho_t^{(exp)}\sim \frac{ m_{s=1} m_{s=1/2}}{m_{s=0}^2} &\simeq & 1.
\end{eqnarray}
Somehow the mass of the ``lowest'' scalar particle mass 
is numerically 
the geometric mean of the highest  spin 1 and spin 1/2 masses.

Let us now assume that the new particle is a scalar Higgs boson. 
In the Standard Model (SM) with a Higgs sector consistent of one
Higgs doublet $\Phi$ and scalar potential 
\begin{eqnarray}
V_{SM}&=& \mu^2\Phi^\dagger \Phi + \lambda \left(\Phi^\dagger \Phi\right)^2,
\end{eqnarray}
all masses are induced by  spontaneous symmetry breaking and are 
proportional to the Higgs vaccuum expectation
$<\phi_0>=v/\sqrt{2}$
where 
\begin{eqnarray}
v^2 &=& -\frac{\mu^2}{\lambda}=\frac{1}{\surd 2 G_F}= (246.218)^2 
\quad \left(\gev\right)^2.
\end{eqnarray}
The tree level top, gauge and Higgs boson masses are given in terms of $v$
and their respective Yukawa couplings
\begin{eqnarray}
m_W &=& g \frac{v}{2}, \quad m_Z=\sqrt{g^2+g'^2} \frac{v}{2} \\
m_t &=& g_t \frac{v}{2},\\
m_H^2 &=& -2 \mu^2= 2 \lambda  v^2. 
\end{eqnarray}
Moreover we have 
$g'=g \tan\theta_W$ or 
$\sqrt{g^2+g'^2}=g/\cos\theta_W$,$ G_F m_W^2/\surd 2=g^2/8$.

In terms of these quantities the tree level mass 
ratio $\rho_t$ is 
simply given by
\begin{eqnarray}
\rho_t^{0(SM)} &=&  \frac{m_Z m_t}{m_H^2}= \sqrt{g^2+g'^2} \frac{v^2 g_t}{4\surd 2 v^2 \lambda} \\
&=& \frac{1}{4\surd 2}\frac{\sqrt{g^2+g'^2} g_t}{ \lambda} \label{eq19}\label{eq22}\label{eqr1}\\
&=& \frac{1}{4\surd 2}\frac{g g_t}{\cos\theta_W \lambda}. 
\end{eqnarray}

Similarly, the tree level SM $\rho_{Wt}$ ratio is given by:
\begin{eqnarray}
\rho_{Wt}^{0(SM)} &=&  \frac{m_W +m_t}{2m_H}= 
 \frac{g+ g_t}{4\surd (2  \lambda)}.
\label{eq19zz}
\end{eqnarray}

Numerically, we estimate the top yukawa, quartic coupling and 
other related ratios 
as ( from measured masses, using 
 tree level approximation, $1\sigma$ errors ) :
{\small
\begin{eqnarray}
g_t^{0,(exp)}&=& 1.409\pm 0.007,\\
\lambda^{0,(exp)} &=& 0.130\pm 0.001,\\
\left (\frac{\lambda}{g_t^2}\right )^{0,(exp)} &=&\frac{1}{8} 
\left(\frac{m_H}{m_t}\right)^2=(6.6\pm 0.1)\times 10^{-2},\label{eq25}\\
%%\left (\frac{\lambda^2}{g_t^2}\right )^{(exp)} &=&0.006175\pm xxx.\\
\left (\frac{g^2+g'{}^2}{\lambda}\right )^{0,(exp)} &=&8 
\left(\frac{m_Z}{m_H}\right)^2=4.21\pm 0.03. \label{eq27b}
\end{eqnarray}
}

%In the SM, the Higgs selfcoupling $\lambda$ is non-determined. However, assuming that the value $\rho_t^{(exp)}\simeq 1$ is not merely a coincidence, the relation Eq.(\ref{eq19}) tell us that the scalar self-coupling and the spin 1 and spin 1/2 top couplings are  subject to the tree level equality
%\begin{eqnarray}\lambda &\simeq& c \sqrt{g^2+{g'}{}^2} g_t \label{eq25b}\\        &\simeq& c g g_t,\end{eqnarray}
%where $c$ is a numeric factor of order $\sim o(1)$.

%%%%%
In the SM, the Higgs selfcoupling $\lambda$ is non-determined. 
However, assuming that 
 \emph{both} expressions 
$\rho_t,\rho_{Wt}\simeq 1$ are not a coincidence, 
the relations (\ref{eq19}) and (\ref{eq19zz}) tell us 
that the scalar self-couplings, gauge couplings and Yukawa 
top couplings are related by two expressions. 
At tree level any two of the quantities $\lambda,g,g',g_t$ 
can be written in terms the two others using the expressions:
\begin{eqnarray}
\lambda &\simeq& c \sqrt{g^2+{g'}{}^2} g_t ,\label{eq25b}\\
\lambda &\simeq& c^2 (g+g_t)^2 \label{eq26bzz}
\end{eqnarray}
where $c$ is a  numeric factor of order $\sim o(1)$.
If we take into account only the first expression, the 
one related to the ratio $\rho_t$, we arrive to 
the relation between the quartic, gauge and top 
couplings 
\begin{eqnarray}
\lambda&\sim& g g_t.
\label{eq29zz}
\end{eqnarray}
Let us take now into account  both expressions. 
For $g_t>>g$ the second equation becomes
 $\lambda \simeq c^2 g_t^2$, inserting it in the first 
one we arrive to
\begin{eqnarray}
\lambda &\simeq& \kappa \left (g^2+{g'}{}^2\right )
\label{eq39zz}
\end{eqnarray}
with $\kappa\simeq 1+o(g/g_t)$.

%%%%%%%%%%%%%%%%%%%%%%
%%%%%%%%%%%%%%%%%%%%%%%%%%%

\vspace{0.3cm}
%{\bf 3.} 
The tree level relations (\ref{eq19},\ref{eq19zz}) are  
affected by SM quantum corrections. 
Including  one loop corrections, the three level relations 
above 
should be replaced, in particular by 
(where $\mu_0$ the renormalization scale, $\mu_0\sim m_Z-m_t$)
\begin{eqnarray}
g_t (\mu_0) &=& \frac{\surd 2 m_t}{v} \left ( 1+\delta_t(\mu_0)\right ),\label{eq27}\\
\lambda (\mu_0) &=& \frac{\surd  m_H^2}{2v^2} \left ( 1+\delta_\lambda(\mu_0)\right ),\label{eq28}
\end{eqnarray}
we consider negligible the running of the gauging couplings
$g_i(\mu_0)$. The first order corrected ratio $\rho_t$ is then,
using expresions (\ref{eq27},\ref{eq28}), 
\begin{eqnarray}
\rho_t^{SM} &=&  \frac{m_Z m_t}{m_H^2}\\
&=& \frac{1}{4\surd 2}\frac{g g_t}{\cos\theta_W \lambda}
\frac{ 1+\delta_\lambda} { 1+\delta_t} \\
&\simeq& \rho_t^0 \left( 1+\delta_\lambda-\delta_t\right).
\label{eqr2}
\end{eqnarray}
The top yukawa $\delta_t$ can be written as 
$\delta_t=\delta_t^{QCD}+\delta_t^w$. 
The corrections are (\cite{refsmcorrections} and references therein), ignoring logarithm terms,
\begin{eqnarray}
\delta_\lambda &=& \frac{1}{16 \pi^2} c_\lambda \lambda,\\
\delta_t^w &=&  \frac{1}{16 \pi^2} \frac{c_t}{8}  g_t^2,\\
\delta_t^{QCD}&=& (-1/(3 \pi^2))  g_s^2, 
\end{eqnarray}
with the numerical coefficients
$c_\lambda \simeq 25/2-9 \pi/(2\surd 3)\simeq 4.3$,
$c_t \simeq 6.1$. Thus
\begin{eqnarray}
\frac{\delta_\lambda}{\delta_t^w} &\simeq & \frac{c_\lambda}{c_t} \left (\frac{m_H}{m_t}\right)^2\simeq 0.3.
\end{eqnarray}
Then
\begin{eqnarray}
\rho_t &=& \rho_t^0\left ( 1+c_1 \lambda-c_2 g_t^2-c_s g_s^2\right ).
\end{eqnarray}
The correction $\delta_t^{QCD}\sim 5\%$ is the most important one, 
acting to diminish slightly the ratio.
Both corrections, $\delta_t^w,\delta_\lambda$, are of opposite sign and very small, of the order of $1\%$.

%%Use the corrections to the inverse, to extract the g g g ratio from the masses...

\section{SM Renormalization group equations.}

We explore here the behaviour of the mass ratio (\ref{eq2},\ref{eq22}) 
at higher scales. 
%%The renormalization group equations up to two loops were 
%%calculated for the different couplings 
%%\cite{wimmer79,wimmer80,wimmer55}. 
%%%
%%The Yukawa couplings of the first and second generation quarks and leptons are so small that lead to negligible contributions for our purposes and they are assumed to vanish in the following.
We consider first a reduced system of one-loop renormalization 
group equations 
where only 
the Higgs self-coupling $\lambda$ and the top Yukawa coupling $g_t$
appear. All the other couplings are considered very small or not 
running at all. The RGE equations for the individual couplings 
take the form  (see for example \cite{wimmer79,wimmer55,wimmer,wimmer40}) (with 
$t=\log(\mu/\Lambda)$ , expression valid for high, but no so high, 
scales $\mu>> m_t,m_H$, or for $\Lambda\to \infty$):
\begin{eqnarray}
\frac{d g_t^2}{d t} &=& \frac{9}{16 \pi^2} g_t^4,\label{eq40}\\
\frac{d \lambda}{d t} &=& \frac{6}{16 \pi^2}\left (4\lambda^2+2 \lambda g_t^2- g_t^4\right).
\end{eqnarray}
If we introduce the variable
\begin{eqnarray}
R&=& \frac{\lambda}{g_t^2},%\quad S= \frac{\lambda^2}{g_t^2},\quad 
\end{eqnarray}
the RGE equations for $g_t,R$ and $\rho_t(t)$ 
become decoupled with nested 
solutions, $g_t=g_t(\mu)$,$ R=R(g_t)$,$\rho_t=\rho_t(R)$.
In addition to Eq.(\ref{eq40}), we have
\begin{eqnarray}
g_t^2\frac{d R}{d g_t^2} &=& \frac{1}{3} f(R),
\label{eq119}\label{eq43}\\
%%g_t^2\frac{d \rho_t}{d g_t^2} &=& -\frac{\rho_t}{2} \left (1+\frac{2 f(R)}{3 R} \right ),\\
\frac{d \rho_t}{d R} &=& -\frac{3\rho_t}{2 f(R)} \left (1+\frac{2 f(R)}{3 R} \right ).\label{eq45}
\end{eqnarray}
with $f(R)=8R^2+R-2$. 
The  equations (\ref{eq40},\ref{eq43},\ref{eq45})
can be solved explicitly, in particular for 
the ratio $\rho_t$ we can write 
$$\rho_t=k \left (\frac{R_0-R}{R_1+R}\right)^{R_0-R_1} R^2,$$
where $R_0,R_1$ are the  fixed points of the  equation (\ref{eq43}), 
$f(R_{0,1})=0$.
For a light Higgs and large top mass the ratio $R$ is small, at low 
scales $R^{exp}\sim  10^{-1}$, Eq.(\ref{eq25}). 
For such a small $R$ the solution
of the differential equations is approximately:
\begin{eqnarray}
R(g_t) &=& R_c-\frac{4}{3}\log g_t,
\end{eqnarray}
and
\begin{eqnarray}
\rho_t&\sim& k R^2
\sim (R_c-\frac{4}{3}\log g_t)^2\sim k R_c^2 \sim \rho_t^0.
\end{eqnarray}
At large energies ($\mu>> m_t$, as long as $R>0$ or $\lambda>0$), 
the ratio $\rho_t(\mu)$ keeps approximately constant, 
only sligtly decreasing with the logarithm of $g_t$.

%%%%%%%%%%%%%%%%%%%%%%%%%%%%%%%%%%%%%%%%%%%%%%%%%%%%%%%
If we consider a reduced
Higgs-top-strong system where the $\lambda,g_t,g_s$ are non-vanishing
and allowed to run together with the ratios $R,\rho_t$. 
One ends with a similar system of equations where the evolution 
of $\rho_t$ is of the type $ g_t^2d \rho_t/g_t^2\sim \rho_t h(R, g_t^2)$
and similar results are obtained.
%%%%%%%%%%%%%%%%%%%%%%%%%%%%%%%%%%%%%%%%%%%%%%%%%%%%%

At higher energies, and for more quantitative results, a 
full treatment is needed. Present state-of-the-art NLO and NNLo 
constraints on SM vacuum stability \cite{anton} seems to guarantee stability, and then a reasonably stable, positive, value for the quartic coupling, for a Higgs mass $m_H\sim 126$ \gev and
to very high scales.. 
If we assume a stable behaviour for $\lambda$ and ignoring the very 
modest variation of the coupling factor $g^2+g'{}^2$, 
$$\rho_t(\mu)\sim \rho_t^0\frac{g_t(\mu)}{g_t^0}.$$
the variation of the mass ratio $\rho_t$ is governed by the 
top Yukawa up to scales where new physics is expected to emerge.

\section{Conclusions and further discussion}

%%6.-
We expect new physics that cuts off the divergent top, 
gauge and 
higgs loop contributions to the Higgs Mass at scales $\lsim 10$ TeV. 
Many different possibilities have been well explored, they usually 
include, more or less ad-hoc, new particles with properties tightly 
associated to those of the SM. Some of these possibilities 
are for example (and any  combinations among them)\cite{arkani,pomarol}:
a) The new particles are just the, 
 softly broken, SUSY, superpartners with couplings and Yukawas 
strongly dictated  by supersymmetry and the  soft breaking itself.
b) The Higgs is a composite resonance, or 
c) The ``Little'' Higgs is a pseudo-Nambu-Goldstone boson with respect 
 a ``softly'' broken  approximate global symmetry. This scalar 
sector is  accompanied by some new  particles belonging to enlarged 
multiplets together with the SM particles.

It is a general feature that,  in all or most of these models,
the quartic self coupling, and then the Higgs mass, is related 
to the gauge coupling constants and to the top yukawa in a more 
or less explicit way, reminding of the relation (\ref{eq25}) suggested by the experimental evidence $\rho_t\simeq 1$. 
The reason is clear \cite{arkani}, 
the new one-loop which are proportional to the couplings of the 
SM gauge sector (or to a subsector of an enlarged gauge sector) 
have to match 
and cancel the top and the other cuadratic loops.

We will briefly review the situation in the MSSM and Littlest Higgs
scenarios.
In the MSSM,
the tree level top, gauge and lowest Higgs boson masses together  
 their respective Yukawa couplings are given by the expressions
\begin{eqnarray}
v^2 &=& v_1^2+v_2^2,\quad \tan\beta=v_2/v_1,\\
m_W &=& g \frac{v}{2}, \quad m_Z=\sqrt{g^2+g'^2} \frac{v}{2} \\
m_t &=& g_t \frac{v}{2}\sin\beta.
%\\m_H^2{}_{tree} &=& -2 \mu^2= 2 \lambda  v^2. 
\end{eqnarray}
where the tree level Higgs quartic coupling is determined in terms 
of the gauge couplings
\begin{eqnarray}
\lambda^0 &=&  (g^2+g'{}^2)\cos^22\beta.
\end{eqnarray}

Quantum corrections to the Higgs quartic coupling 
 are very important. They lead for an expression for the 
lower neutral  Higgs mass, of the form \cite{blum62}
\begin{eqnarray}
m_H^2 &=& m_Z^2 \cos^2 2\beta+ \delta m_H^2 \\
     &=& m_Z^2 \cos^2 2\beta+ f \frac{3 G_F}{\sqrt{2}\pi^2} m_t^4 
\label{mhmssm} 
\end{eqnarray}
where the factor $f$ include logarithmic corrections, mainly 
associated to stops. From the 
expression (\ref{mhmssm}) and from  $m_H^2= 2 \lambda(\mu) v^2$ we 
can extract an improved value for the quartic effective coupling
\begin{eqnarray}
\lambda (\mu) &=& \frac{m_H^2}{2 v^2}\left (1+\delta_\lambda(\mu)\right ).
\end{eqnarray}
The effective quartic coupling is of the natural size $\lambda\sim 
o(g^2, g_t^4)$.
In terms of these quantities the  mass ratio $\rho_t$ is 
simply given by
\begin{eqnarray}
\rho_t^{MSSM} &=&  \frac{m_Z m_t}{m_H^2}=  \\
&=& \frac{\sqrt{g^2+g'^2} g_t \sin\beta}{(g^2+g'^2) \cos^22\beta+ g_t^4 \sin^4\beta 3f/\pi^2}.
%,\\ &=& \frac{\sqrt{g^2+g'^2} g_t \sin\beta}{(g^2+g'^2) \cos^22\beta+ g_t^4 \sin^4\beta 3f/\pi^2}. 
\end{eqnarray}
In the context of the MSSM, 
the experimental evidence $\rho_t\simeq 1$ is a strong hint for the 
existence of a relation among the parameters of the expression above,
principally  top Yukawa and $\tan\beta$ together with the gauge couplings.

As a second illustrative example, let us mention
 the ``Littlest'' Higgs scenario
\cite{arkani}. Here the usual Higgs doublet, is the lightest 
 of a set of pseudo goldstone bosons in an non-linear sigma model 
including in its gauge group different $SU(2)\times U(1)$ factors.
The product group is broken to the diagonal, identified as the 
SM electroweak gauge group.
The top Yukawa coupling generates 
a negative mass squared triggering electrowak symmetry breaking.
New particles are added, in particular heavy top partners, 
which cancel 
the one loop quadratically divergent corrections.
The quartic self coupling is related to the coupling constants of the
gauge sector and to the top Yukawa with a natural size 
$$\lambda\sim o(g^2,g_t^2),$$ 
reminding, for example, the expression (\ref{eq29zz})
suggested by experimental evidence. % $\lambda\sim g g_t$.
Particular scenarios can be tuned so that either 
the gauge contributions or the top Yukawas 
dominate the Higgs quartic and $m_H\sim m_Z$ or $m_H\sim m_t$ as 
extreme cases.
In fact we have seen, according to Eq.(\ref{eq2}),  that nature chooses, to a very high precision, just
the geometric average of both extreme cases $m_H=\sqrt{m_Z m_t}$.
It seems plausible that a Little Higgs scenario can be found where 
this value appears naturally.
Approximate accidental global 
symmetries related to the Little Higgs scenario could play a 
role in the understanding  of the $\rho_t$ ratio, as the global 
custodial $SU(2)_c$ symmetry \cite{custodial} plays for the $\rho$ ratio.

%\section{Summary and conclusions}

%%%%

In this short note we have presented some simple computations 
associated to  the ratio of the product of $Z^0$ and top masses 
to the squared Higgs mass, $\rho_t$. 
We have shown how this ratio is suprisingly and robustly 
close to the unity at the $10^{-3}$ level. 
The Higgs mass seems to be just the geometrical 
mean of the $m_Z$ and $m_t$ masses.

We have briefly reviewed the theoretical predictions of this ratio 
in the SM and  beyond. 
In the SM, the Higgs selfcoupling $\lambda$ is undetermined. 
However, assuming that the value 
$\rho_t^{(exp)}\simeq 1$ %, $\rho_{Wt}^{(exp)}\simeq 1$ 
is not 
merely a coincidence, the relation Eq.(\ref{eq19}) tell us that 
the scalar self-coupling and the spin 1 and spin 1/2 top couplings 
are 
 subject to the tree level equality
\begin{eqnarray}
\lambda &\simeq& c \sqrt{g^2+{g'}{}^2} g_t \simeq c g g_t,
\end{eqnarray}
where $c$ is a numeric factor of order $\sim o(1)$. Such a relation 
is not very much affected by quantum effects at least up to scales 
$\mu\sim m_t$ or slightly higher. 

%%% 
Moreover, the combined use of both the relations
$\rho_t^{(exp)}\simeq 1$ , $\rho_{Wt}^{(exp)}\simeq 1$ 
leads to  the SM tree level relation
(not very much affected by quantum effects) 
\begin{eqnarray}
\lambda &\simeq& g^2+{g'}{}^2. 
\end{eqnarray}

In a model independent way, 
the   relations $\rho_t\simeq\rho_{Wt}\simeq 1$ imply that 
any two of the quantities $m_H,m_W,m_Z,m_t$ can be written
in terms of the other two. Taking into account also 
the relation $\rho\simeq 1$ we can write any two mass 
ratios as a function of, exclusively, the Weinberg angle 
(with a precision of the order of $1\%$ or better)
$\frac{m_i}{m_j}\simeq  f_{ij}(\theta_W)$,$ i,j=W,Z,H,t$.
In the limit $\cos\theta_W\to 1$ all the masses 
would become equal $m_Z=m_W=m_t=m_H$.
The existence of such relations could be interpreted as a hint for an 
instrumental role, together with some other unknown symmetry, 
of the $SU(2)_c$ custodial symmetry in the explanation of
the $m_H/m_t$ ratio \cite{custodial}.

The closeness of the parameter $\rho_t,\rho_{Wt}$ to one, 
``the Higgs mass coincidence problem'',
might be merely a 
coincidence or an artifact of the early status of the 
Higgs discovery, 
which will dissapear with any new measurement.
It is tempting to think that such a  value of  the ratios
 are, on naturalness grounds, a signal of a  deeper 
 mechanism or symmetry.

%%%%%%%%%%%%%%%%%%%%%%%%%%%%%%%%%%%%%%%%%%%%%%%%%
\vspace{.5cm}
\acknowledgments

The author wish to thank to the CERN TH dept. 
for its hospitality during the realization of this work.
This work has been  supported in part by a grant from the
Spanish Ministry of Science.

%%%%%%%%%%%%%%%%%%%%%%%%%%%%%%%%%%%%%%%%%%%%%%%%%%%%%%%%%%%%%%%%%%


\begin{thebibliography}{99}








%%%%%%%%%%%%%%%%%%%%%%%%%%%%%%%%%%%%%%%%%%%%%%%%%%%%%
%%%%%%%%%%%%%%%%%%%%%%%%%%%%%%%%%%%%%%%%%%%%%%%%%%%%%%%%

%\cite{pdg2010}
\bibitem{pdg2012}
%%%      K.~Nakamura  et al. (Particle Data Group), 
%%%   J.\ Phys.\ G {\bf 37}, 075021 (2010).
%The Review of Particle Physics.
 J. Beringer et al. (Particle Data Group), 
Phys. Rev. D86, 010001 (2012)
 and 2013 partial update for the 2014 edition. 

\bibitem{atlas}
%\bibitem{Aad:2012tfa}
  G.~Aad {\it et al.}  [ATLAS Collaboration],
  %``Observation of a new particle in the search for the Standard Model Higgs boson with the ATLAS detector at the LHC,''
  Phys.\ Lett.\ B {\bf 716} (2012) 1
  [arXiv:1207.7214 [hep-ex]].
  %%CITATION = ARXIV:1207.7214;%%
  %1849 citations counted in INSPIRE as of 02 Nov 2013
%\cite{ATLAS:2012oga}
%\bibitem{ATLAS:2012oga}
  G.~Aad {\it et al.}  [ATLAS Collaboration],
  %``A particle consistent with the Higgs Boson observed with the ATLAS Detector at the Large Hadron Collider,''
  Science {\bf 338} (2012) 1576.
  %%CITATION = SCIEA,338,1576;%%
  %19 citations counted in INSPIRE as of 02 Nov 2013
%\cite{Aad:2013xqa}
%\bibitem{Aad:2013xqa}
  G.~Aad {\it et al.}  [ATLAS Collaboration],
  %``Evidence for the spin-0 nature of the Higgs boson using ATLAS data,''
  Phys.\ Lett.\ B {\bf 726} (2013) 120
  [arXiv:1307.1432 [hep-ex]].
  %%CITATION = ARXIV:1307.1432;%%
  %34 citations counted in INSPIRE as of 02 Nov 2013
%\cite{Aad:2013wqa}
%\bibitem{Aad:2013wqa}
  G.~Aad {\it et al.}  [ATLAS Collaboration],
  %``Measurements of Higgs boson production and couplings in diboson final states with the ATLAS detector at the LHC,''
  Phys.\ Lett.\ B {\bf 726} (2013) 88
  [arXiv:1307.1427 [hep-ex]].
  %%CITATION = ARXIV:1307.1427;%%
  %54 citations counted in INSPIRE as of 02 Nov 2013
%\cite{Orestano:2013dea}
%\bibitem{Orestano:2013dea}
  D.~Orestano [ATLAS Collaboration],
  %``Search for the standard model Higgs boson with the ATLAS detector,''
  Int.\ J.\ Mod.\ Phys.\ D {\bf 22} (2013) 1330015.
  %%CITATION = IMPAE,D22,1330015


\bibitem{cms}
%\bibitem{Chatrchyan:2012ufa}
  S.~Chatrchyan {\it et al.}  [CMS Collaboration],
  %``Observation of a new boson at a mass of 125 GeV with the CMS experiment at the LHC,''
  Phys.\ Lett.\ B {\bf 716} (2012) 30
  [arXiv:1207.7235 [hep-ex]].
  %%CITATION = ARXIV:1207.7235;%%
  %1831 citations counted in INSPIRE as of 02 Nov 2013
%\cite{Chatrchyan:2012jja}
%\bibitem{Chatrchyan:2012jja}
  S.~Chatrchyan {\it et al.}  [CMS Collaboration],
  %``Study of the Mass and Spin-Parity of the Higgs Boson Candidate Via Its Decays to Z Boson Pairs,''
  Phys.\ Rev.\ Lett.\  {\bf 110} (2013) 081803
  [arXiv:1212.6639 [hep-ex]].
  %%CITATION = ARXIV:1212.6639;%%
  %68 citations counted in INSPIRE as of 02 Nov 2013






\bibitem{refsmcorrections}
%\cite{Wetterich:1987az}
%\bibitem{Wetterich:1987az}
  C.~Wetterich,
%  ``The Mass Of The Higgs Particle,''  
DESY-87-154.  
%%CITATION = DESY-87-154;%%
%\cite{Altarelli:1994rb}
%%\bibitem{Altarelli:1994rb}
  G.~Altarelli and G.~Isidori,
  %``Lower limit on the Higgs mass in the standard model: An Update,''
  Phys.\ Lett.\ B {\bf 337} (1994) 141.  
%%CITATION = PHLTA,B337,141;%%
%%\bibitem{otros:fixpoints}
%\cite{Pendleton:1980as}
%\bibitem{Pendleton:1980as}
  B.~Pendleton and G.~G.~Ross,
%  ``Mass and Mixing Angle Predictions from Infrared Fixed Points,'' 
 Phys.\ Lett.\ B {\bf 98} (1981) 291.  
%%CITATION = PHLTA,B98,291;%%


\bibitem{wimmer79}
M.E. Machacek and Vaughn, Nucl. Phys {\bf B222} (1983) 83; ibid. 
{\bf B236} (1984) 221; ibid.{\bf B249} (1985) 70;
C. Ford, D.R.T. Jones, P.W. Stephenson and M.B. Einhorn, Nucl. Phys. 
{\bf B395} (1993) 17.
K. Inoue, A. Kakuto and S. Takeshita, Prog. Theor. Phys. 67 (1982) 1889; ibid. {\bf 68} (1982) 927;
S.P. Martin and M.T. Vaughn, Phys. Rev. {\bf D50} (1994) 2282.
%\cite{Cerdeno:2001se}
%%%\bibitem{varios}
  D.~G.~Cerdeno, E.~Gabrielli, S.~Khalil, C.~Munoz, E.~Torrente-Lujan 
  %``Determination of the string scale in D-brane scenarios and dark matter implications,''  
Nucl.\ Phys.\ B {\bf 603} (2001) 231  [hep-ph/0102270].  
%%CITATION = HEP-PH/0102270;%%
 %\cite{Akers:1994wi}
%%%%%%%%%%%%%%%%%%%%%%%\bibitem{Akers:1994wi}
  R.~Akers {\it et al.},
  %``Search for the minimal Standard Model Higgs boson,''  
Phys.\ Lett.\ B {\bf 327} (1994) 397.  
%%CITATION = PHLTA,B327,397;%%




\bibitem{wimmer55}
%\cite{Barger:1992ac}
%\bibitem{Barger:1992ac}
  V.~D.~Barger, M.~S.~Berger and P.~Ohmann,
  %``Supersymmetric grand unified theories: Two loop evolution of gauge and Yukawa couplings,'' 
 Phys.\ Rev.\ D {\bf 47} (1993) 1093  [hep-ph/9209232].  
%%CITATION = HEP-PH/9209232;%%

\bibitem{wimmer}
  B.~Schrempp and M.~Wimmer,
  %``Top quark and Higgs boson masses: Interplay between infrared and ultraviolet physics,'' 
 Prog.\ Part.\ Nucl.\ Phys.\  {\bf 37} (1996) 1  [hep-ph/9606386].  %%CITATION = HEP-PH/9606386;%%



\bibitem{wimmer40}
C. Wetterich, Proc. Trieste HEP Workshop (1987) p.403, and preprint DESY-87-154
(1987).


 


\bibitem{arkani}
%\bibitem{ArkaniHamed:2002qy}
  N.~Arkani-Hamed, A.~G.~Cohen, E.~Katz and A.~E.~Nelson,
  ``The Littlest Higgs,''  
JHEP {\bf 0207} (2002) 034  [hep-ph/0206021].  %%CITATION = HEP-PH/0206021;%%
%\bibitem{little}%\cite{Schmaltz:2005ky}
%\bibitem{Schmaltz:2005ky}
  M.~Schmaltz and D.~Tucker-Smith,
  ``Little Higgs review,''  
Ann.\ Rev.\ Nucl.\ Part.\ Sci.\  {\bf 55} (2005) 229  [hep-ph/0502182].  %%CITATION = HEP-PH/0502182;%%


\bibitem{pomarol}
  A.~Pomarol,
  %``Beyond the Standard Model,''  
CERN Yellow Report CERN-2012-001, 115-151  [arXiv:1202.1391 [hep-ph]].  %%CITATION = ARXIV:1202.1391;%%
 %\cite{Gripaios:2009pe}
%%%\bibitem{Gripaios:2009pe}
  B.~Gripaios, A.~Pomarol, F.~Riva and J.~Serra,
  %``Beyond the Minimal Composite Higgs Model,''  
JHEP {\bf 0904} (2009) 070  [arXiv:0902.1483 [hep-ph]].  
%%CITATION = ARXIV:0902.1483;%%




%\bibitem{Anchordoqui:2012fq}
\bibitem{anton}
  L.~A.~Anchordoqui, I.~Antoniadis, H.~Goldberg, X.~Huang, D.~Lust, T.~R.~Taylor and B.~Vlcek,
  %``Vacuum Stability of Standard Model^{++},'' 
 arXiv:1208.2821 [hep-ph].  
%%CITATION = ARXIV:1208.2821;%%
%%%\bibitem{anton1222}
 M. Lindner, M. Sher and H. W. Zaglauer, Phys. Lett. B{\bf 228}, 139 (1989).
 M. Sher, Phys. Rept. {\bf 179}, 273 (1989).
 J. Ellis, J. R. Espinosa, G. F. Giudice, A. Hoecker and A. Riotto, 
Phys. Lett. B {\bf 679}, 369(2009) [arXiv:0906.0954 [hep-ph]].
 Z. -z. Xing, H. Zhang and S. Zhou, arXiv:1112.3112 [hep-ph].
%%%%\bibitem{anton39}
 L. J. Hall and Y. Nomura, JHEP {\bf 1003}, 076 (2010) [arXiv:0910.2235 [hep-ph]].




%%%%%%
\bibitem{blum62}
M. Drees, R. Godbole and P.Roy, {\em
Theory and phenomenology of sparticles: An account of four-dimensional N=1 supersymmetry in high energy physics.}
Hackensack Ed.,  World Scientific (2004).



% Custiodial symmetry
\bibitem{custodial}
  P.~Sikivie, L.~Susskind, M.~B.~Voloshin and V.~I.~Zakharov,
%  ``Isospin Breaking In Technicolor Models,''
  Nucl.\ Phys.\  B {\bf 173}, 189 (1980).
  %%CITATION = NUPHA,B173,189;%%
% Discussion of different custodial THDMs
%\cite{Pomarol:1993mu}
  A.~Pomarol and R.~Vega,
%  ``Constraints on CP violation in the Higgs sector from the rho parameter,''
  Nucl.\ Phys.\  B {\bf 413}, 3 (1994)
  \mbox{[arXiv:hep-ph/9305272].}
  %%CITATION = NUPHA,B413,3;%%
%\cite{Gerard:2007kn}
  J.~M.~Gerard and M.~Herquet,
  %``A twisted custodial symmetry in the two-Higgs-doublet model,''
  Phys.\ Rev.\ Lett.\  {\bf 98}, 251802 (2007)
  \mbox{[arXiv:hep-ph/0703051].}
  %%CITATION = PRLTA,98,251802;%%

\bibitem{torrente2}
E. Torrente-Lujan,'' The Higgs and top masses, the ratio $\rho_t$ and
the $SU(2)_c$ custodial symmetry `` (To appear).



\end{thebibliography}
\end{document}